**SOME COMPLICATIONS OF THE ELEMENTARY FORMS OF COMPETITION IN A SOURCE/SINK AND METACOMMUNITY CONTEXT: THE ROLE OF INTRANSTIVE LOOPS**

John Vandermeer

Department of Ecology and Evolutionary Biology

University of Michigan

Ann Arbor MI 48109

jvander@umich.edu

Dynamic interspecific interactions of many forms (e.g., parasitism, competition, mutualism) are thought to contribute to fundamental structures of ecological communities.  Of the various types of interactions, the process of ecological competition has loomed large in the literature attempting to give theoretical structure to community ecology (e.g., Elton, 1946; Bray and Curtis 1957; Connell, 1961; Vandermeer, 1969; Pianka , 1973; Cody, 1974; Wiens, 1977; Tilman, 1981; Lawton and Strong, 1981; Schoener 1982; Roughgarden, 1983; Ricklefs, 1987; Simberloff and Dayan, 1991; Welborn et al., 1996; Chesson, 2000; McCann, 2000; Webb, 2000; Chase et al., 2002; Bever, 2003; Amarasekare, 2003; Silvertown, 2004; Hubbell, 2005; Wiens and Graham, 2005; McGill et al., 2006; Mayfield and Levine, 2010). Summarizing the most salient features of competition as a "phenomenon," regardless of what might be its mechanism, has usually been aided by the Lotka Volterra competition equations, in which the competitive process is thought of as a linear term added to the elementary logistic form of population growth, usually presented in its biologically sensible form,

$$\frac{dx_i}{dt} = r_i x_i \left( \frac{K_i - x_i - \sum_{j \neq i} \alpha_{ij} x_j}{K_i} \right),$$
1

where $x_i$ is the biomass or population density of species i, $r_i$ is the intrinsic growth rate of species i, $K_i$ is the carrying capacity of species i and $a_{ij}$ is the competitive effect of species j on species i (normally thought to be the ratio of interspecific to intraspecific competition). From this there emerge four qualitatively distinct outcomes of two-species competition (coexistence, species 1 wins, species 2 wins, or indeterminate, with one or the other winning, depending on founding conditions). Extrapolating the 2D form of these simple equations into a community (multispecies) formulation seems straightforward – simply keep adding the linear terms and consider the multispecies community as that linear combination of competition effects, whence equation 1 is the evident result (e.g., Levins, 1968;

Vandermeer, 1969; Goh, 1977; Fan, et al., 1999; Kokkoris et al., 1999; May, 2001; Jin et al., 2004; Takeuchi, 1996).

The original two-species form led to an idea that sometimes was regarded as a kind of basic principle of community ecology, when two species compete only weakly, they may coexist in the environment, but if they compete too strongly one or the other must win and exclude the other, known as Gause's principle (the competitive exclusion principle). Extrapolating this principle to larger communities, many authors have effectively suggested that the principle could be applied in a larger context (e.g., Levins, 1968; Levin, 1970; Darlington, 1972; Silvertown, 2004). For example, in one forum of professional ecologists (the 1944 meetings of the British Ecological Society) some of the world's most highly respected ecologists formally debated whether Gause's principle should be taken as a foundational principle for all of ecology (Gilbert et al., 1952). Others have suggested less ambitious agendas for generalization. One ingenious framework is Yodzis' triple classification of founder-controlled, dominance-controlled, and niche-controlled communities (Yodzis, 1978; 1989). Founder-controlled refers to the indeterminate case, in which the ultimate winner in competition depends on initial conditions, where the "founding" species will likely win the competitive struggle. Dominance-controlled refers to those cases in which one species dominates the other, in a pair-wise comparison. The dominance-controlled category includes two distinct forms, a "stable" form and an "unstable" form. The "niche-controlled" category accounts for those communities characterized by weak competition (effectively likening competition to niche overlap, a common shorthand in theoretical ecology, where small niche overlap suggests weak competition). This "periodic table" of competitive effects can be conveniently summarized by phase diagrams with zero growth isoclines (Figure 1).

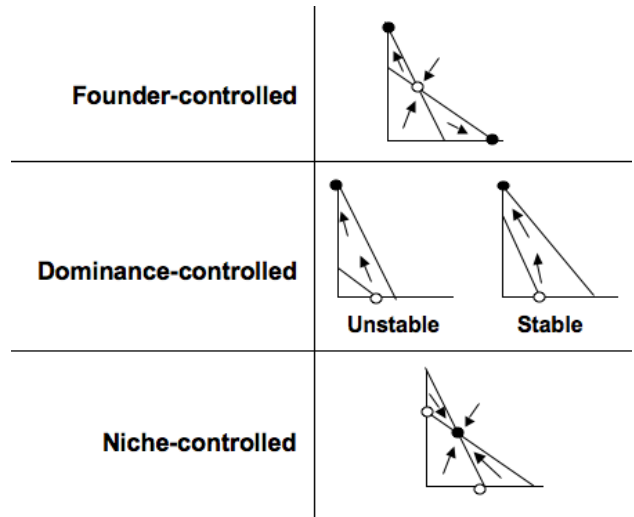

*Figure 1.   Elementary forms of the classical Lotka-Volterrra Equations, with classificatory schemes according to Yodzis.*

In more recent literature the idea of a metacommunity has become popular (Wilson, 1992; Holt, 1993; Hubbell, 2001; Mouquet and Loreau, 2003; Leibold et al., 2004; Holyoak, 2005; Chase, 2005; Urban et al., 2008), to some extent an outgrowth of the idea of collections of source/sink populations, frequently referred to as island/mainland or source/sink communities (Holt, 1993; Mouquet and Loreau, 2002; Rex et al., 2005).  MacArthur and Wilson's island/mainland framework is especially relevant in that it represented a change in focus for population and community studies in general, as eloquently summarized by Hubbell (2001).  Much in the same way that a source/sink population becomes a metapopulation when the source is eliminated (or is too far from the collection of sinks) (see fig 2a), we can imagine an island/mainland community becoming a metacommunity when the mainland is too far removed (see fig 2b).

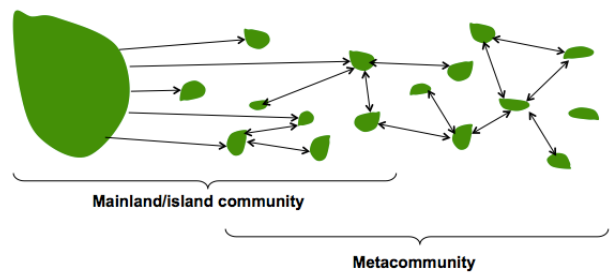

*Figure 2.  Conceptualization of mainland/island community and metacommunity as points along a gradient. Large area symbolizes mainland and small areas islands, or subpopulations.*

Here I conceive of an island/mainland community in the same framework as MacArthur and Wilson's original offering, in which ecological dynamics occur locally, with species interactions of various forms determining which species will survive and which will perish, while the more regional process of migration continually feeds these local communities.  Thus we conceive of a general species pool on the mainland, feeding species to the isolated communities on the islands.  It is likely that these far removed islands also feed one another with migrants in the same way that those migrants come from the mainland.  Islands far enough away from the mainland, if migration is allowed between them but restricted from the far off mainland, become collectively a metacommunity. Eschewing some recent complexities, I thus consider a metacommunity as structured in the original sense of Wilson (1992).  Ecological dynamics (here restricted to competition) occur at a local level, but local patches communicate with one another through dispersal.

In the case of a true island/mainland community, the dynamics of a single local patch (island) are continually repopulated by migrations from the mainland, and, most importantly, from among one another.  Thus the species pool is the collection of species that exist in perpetuity on the mainland, and each island can be expected to house a subset of those species.  In the case of a metacommunity (i.e., no mainland), if we presume that the metacommuity is globally stable, which is to say, all species at the regional level persist in perpetuity, the dynamics of the metacommunity can be studied at the level of a single local patch, and a species pool (the species existing in the metacommunity as a whole).  The question then becomes, how does the Yodzis classification inform the dynamic processes expected to operate in a metacommunity or mainland/island context.

In what follows I first discuss the relevance of adding the stable versus unstable categories to dominance-controlled communities, and second discuss the expected outcome at a local level of a community composed of dominance controlled species pairs in the overall species pool (either the overall metacommunity pool or the mainland pool), explicitly examining the role of intransitive loops in community structure.  In the case of a metacommunity, I assume a stable situation, in the sense that the overall species list for the union of all sub communities is constant.  That is, there may be dramatic changes in number and composition of species in any given subcommunity, but the collection as a whole retains all species.  The collection as a whole, then, can simply be thought of as a species pool, much as a source in a source/sink population or the "mainland" in MacArthur and Wilson's island biogeography.  What can be said about the structure of a local community?  It seems an obvious question since most observations from the real world are effectively made at this level.  It is broadly assumed that this local structure will be due to something like an immigration rate (from the species pool) minus an extinction rate.  It is the extinction process that I seek to explore here, using classical ecological models and assuming that competitive outcome drives that local extinction.

**Stability in dominance-controlled communities**: That there are two forms of dominance is not at all surprising, since it is standard knowledge that if the carrying capacities are equal, the determinate of the community matrix (the matrix of alphas) dictates the stability of any point that exists in the positive quadrant.  That is, if $DetA = 1-\alpha_{ij}\alpha_{ji} > 0$, the system is stable (a node), otherwise it is unstable (a saddle point).  However, with dominance as defined in figure 1, the stability question, while mathematically the same, is somewhat more complicated. In this case, species 1 may be destined to win if either $DetA > 0$ or $DetA < 0$, which is to say the dominant species (the one that wins) will win no matter what the stability, as defined by the competition coefficients.  For the qualitative generalizations traditionally thought important from the LV equations, this does not matter (i.e., for the four distinct outcomes of competition).  But there are several formulations for which it does matter, depending on the question posed.

For example, we might ask what will be the consequence in a two-species dynamic system if equations 1 apply and we slowly increase the value of one of the carrying capacities in the dominance-controlled situation. If the system is stable we have a gradual change in the equilibrium value of the species for which the carrying capacity is being changed. However, if the system is unstable, we have a catastrophic change at the critical point, $K_2 = K_1/a_{12}$, before which the system is completely dominated by $x_1$ and after which it has three equilibria, two stable and one unstable. Then there is another critical point at, $K_1 = K_2/a_{21}$ after which the system is completely dominated by $x_1$. Catastrophic shifts occur at both of those critical points and there is a hysteretic zone between the two. This general result is illustrated in figure 3.

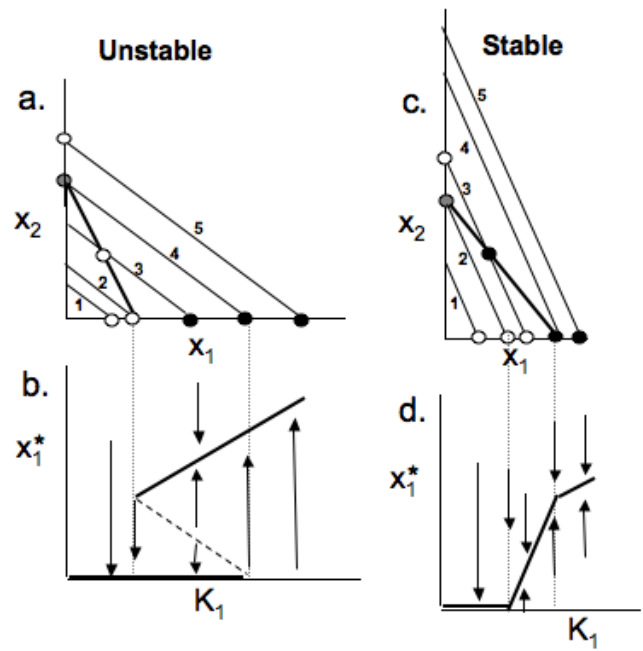

*Figure 3. Response curves to changes in the carrying capacity of species 1. a. the unstable dominance controlled situation, where 5 distinct values of $K_1$ are indicated along with the isocline and equilibrium points (open = unstable; black = stable, grey either unstable or stable depending on which isocline. b. the equilibrium values as a function of the carrying capacity of species 1, illustrating the catastrophic transition from $x_2$ domination to $x_1$ domination. c. the stable dominance controlled situation, where 5 distinct values of $K_1$ are indicated along with the isocline and equilibrium points. d. the equilibrium values as a function of carrying capacity of species 1, illustrating a simple monotonic response.*

**Persistence in dominance-controlled metacommuities**: When communities are fractionated such that the competitive process occurs locally but is fed by regional migrations (i.e., either a mainland/island community or a metacommnity), the long term expectations are not necessarily what is suggested by a direct neighborhood stability analysis. While the expectations from a panmictic community are obvious (e.g., survival of only one species when all competition coefficients are large or persistence of many species when all competition coefficients are all small, i.e., a niche-controlled community,), the case of dominance-control in a metacommunity may lead to other complications. In particular, intransitive loops may arise in which case the pairwise stability question is key to understanding permanence of the species in the loop.

Assume that the community structure is such that all species in the species pool sort out such that every pair of species is dominance-controlled (a simple example would be a collection of species in a strict dominance hierarchy). At one extreme we assume that the apportionment of competition coefficients in the overall species pool is random, yet the structure is organized in such a way that every species pair is dominance-controlled. Then, for a random sampling of species triplets there will be many cases (precisely 25%) in which an intransitive loop will result. It is elementary to demonstrate that an intransitive loop will exist if all of the following three conditions are true:

$$\frac{K_2}{\alpha_{21}} < K_1 < \frac{K_3}{\alpha_{31}} \qquad\qquad 2a$$

$$\frac{K_3}{\alpha_{32}} < K_2 < \frac{K_1}{\alpha_{12}} \qquad\qquad 2b$$

$$\frac{K_1}{\alpha_{13}} < K_3 < \frac{K_2}{\alpha_{23}} \qquad\qquad 2c$$

or if an equivalent set of relations with all inequalities reversed is true. Stability conditions for each pair of species derives directly from the elementary LV equations, namely, the 2D system will be stable if

$$\alpha_{ij}\alpha_{ji} < 1, \qquad\qquad\qquad\qquad\qquad\qquad 3$$

otherwise unstable (in the sense of figure 1). Thus, equations 2 determine the existence of an intransive loop and equation 3 establishes the stability of each of the species pairs involved in the intransitive loop. Combining various combinations of dominance-controlled pairs comprising a three-species subpopulation, generates some perhaps surprising results.

Consider first a three species intransitive loop of dominance-controlled species pairs, where each of those pairs is unstable (i.e. condition 3 is violated). This situation leads to a heteroclinic cycle focused on the three equilibrium points $\{K_1, 0, 0\}$, $(\{0, K_2, 0\}$, and $\{0, 0, K_3\}$ (Vandermeer, 2011; Shi, et al., 2010 ), effectively resulting in a single surviving species (since the heteroclinic cycle gets ever closer to each of the successive points, leading eventually to the stochastic local extinction of one of the species, leaving the dominance structure to eliminate the remaining two species).

The ecological significance of a heteroclinic cycle has been discussed repeatedly (May and Leonard, 1975; Vandermeer, 2011; Schreiber and Killingback, 2013; Hofbauer and Sigmund, 1998; Huisman and Weissing, 2001). Basically, three points are approached repeatedly in an oscillatory sequence of species 1 dominant followed by species 2 dominant followed by species 3 dominant, mathematically approaching the three points in the limit. Yet there is some, even if very small, population density where the population in question must be judged extinct, thus leading to the elimination of two of the three species. Thus, the heteroclinic cycle is ecologically equivalent to Yodzis' founder-controlled community, even though the three component parts are strictly dominance controlled! That is, the species with the initial advantage, most abundant at the "founding" of the community, will generally dominate the other two species

The alternative extreme, when each pair of species in the intransitive loop is Lotka-Volterra stable (LVS), but dominance-controlled (i.e., one species will win in head to head competition, as in figure 1b), a single focal point may be established with all species persistent (as discussed below, this result is not inevitable). This result is somewhat surprising since all three species pairs are themselves non-persistent (i.e., one of the species will dominate the other, in head-to-head pairwise competition), yet when placed together, they may form a stable oscillatory triad (Vandermeer, 2011). However, the formal stability of the equilibrium point is not simply the reverse of equation 3. In particular, if we allow for the system to be symmetrical, the qualitative distinction between stable and unstable can be precisely stipulated. Normalizing to $K_i = 1$ for all i, the equilibrium form of equation 1 (for three species) can be written as,

$$\begin{vmatrix} 0 \\ 0 \\ 0 \end{vmatrix} = \begin{vmatrix} 1 & \alpha_{12} & \alpha_{13} \\ \alpha_{21} & 1 & \alpha_{23} \\ \alpha_{31} & \alpha_{23} & 1 \end{vmatrix} \begin{vmatrix} x_1 \\ x_2 \\ x_3 \end{vmatrix}.$$

Application of equations 2 we see that $a_{21}$, $a_{32}$, $a_{13} > 1.0$, and $a_{31}$, $a_{12}$, $a_{23} < 1$, stipulate an intransitive loop. Consider the special case in which $A = a_{21} = a_{32} = a_{13}$ and $B = a_{31}$, $= a_{12} = a_{23}$, such that the detached coefficient matrix becomes,

$$\begin{vmatrix} 1 & \alpha_{12} & \alpha_{13} \\ \alpha_{21} & 1 & \alpha_{23} \\ \alpha_{31} & \alpha_{32} & 1 \end{vmatrix} = \begin{vmatrix} 1 & B & A \\ A & 1 & B \\ B & A & 1 \end{vmatrix}. \qquad\qquad 4$$

It has been established that,

$$A + B > 2 \qquad\qquad 5$$

stipulates that the equilibrium point is unstable, while the reverse inequality stipulates stability (perfect equality indicates a Hopf bifurcation) (Roeger, 2004; Chi et al., 1998). The subset of conditions implied by equation 3 become

$$AB < 1 \qquad\qquad\qquad\qquad\qquad\qquad 6$$

Equations 5 and 6, although formally applying to only the symmetrical case (i.e., A = $a_{21} = a_{32} = a_{13}$ and B = $a_{31}$, = $a_{12} = a_{23}$ ), give insight into the expected structure on islands in a mainland island framework or in subcommunities in a metacommunity, as illustrated in figure 4.

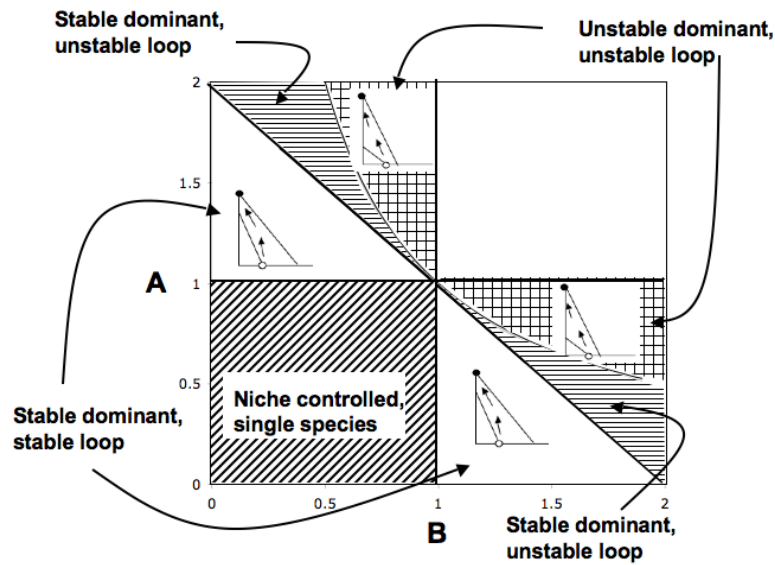

*Figure 4. Generalized structure of stability of the symmetrical system, A versus B (see equation 4), illustrating the role of the stability of the constituent 2D subsystems comprising the community. Stable or unstable dominant refer to the situation of each species pair forming the triplet. Thus if the triplet consists of species A, B and C, the pairs A,B; A,C; and B,C are all characterized by dominance in the Yodzis sense, although that dominance can be either stable or unstable (see figure 1).*

While the conditions of stability are evident in the symmetrical case, relaxing the symmetrical assumption suggests the basic pattern of the symmetrical case is only quantitatively distorted. Generally, if any one of the pairs involved in the intransitive loop is stable, the intransitive loop leads to a focal point attractor of the three species system, albeit dramatically skewed towards one of the species. There is a qualitative distinction that appears between intransitive loops constructed of three stable pairs and those that include only one or two stable pairs (although the formal result of a stable focus is consistent). In figure 5 we illustrate the three qualitatively distinct outcomes.

These results seem to challenge the Yozis categorizations. If it is true that a three species combination for which all two-species subsets are dominance controlled can form either a stable oscillatory equilibrium or a heteroclinic cycle, the actual behavior of the three-dimensional system will be either founder controlled (when there is a heteroclinic cycle) or niche controlled (when there is a stable focal point). So even if the underlying species pool consists of species which, when systematically paired with all other species in the pool, always results in a dominance controlled two-species structure, if the resulting subcommunity forms an intransitive loop (which is expected by chance in 25% of random samples), that subcommunity must be categorized as either founder controlled or niche controlled within the Yodzis framework.

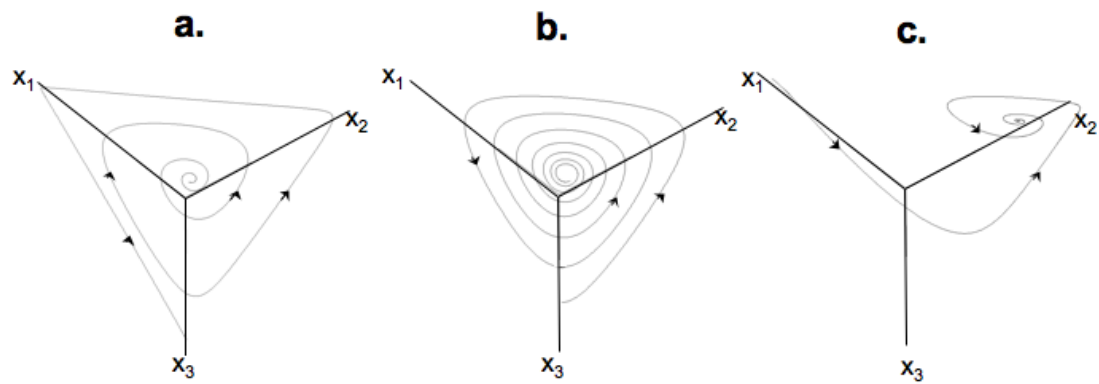

*Figure 4. The three qualitatively distinct outcomes of a three species intransitive loop, formed by three pairs of species, each of which is dominance-controlled. a. when all two-species subsystems are unstable, the result is a heteroclinic cycle.. b. when all two-species subsystems are stable, the result is a symmetrical focal point. c. when one or two of the three two-species subsystems are stable, the result is an asymmetrical focal point.*

**Adding species to an intransitive loop**: Presuming that the first three species arriving in the subcommunity form an intransitive loop, and presuming that the loop is of the stable variety, it is most natural to ask what might happen when a fourth species is added to the system. Presuming the basic symmetrical form of equation 4, and keeping the symmetry when adding the fourth species (i.e., it must form a dominance controlled pairwise relationship with all of the three species in the intransitive loop), the fourth species can be added in four qualitatively distinct forms (figure 5).

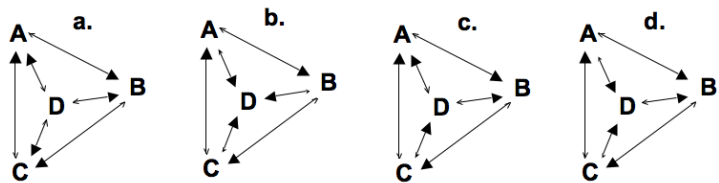

*Figure 5. Various forms of adding a fourth species to a basic intransitive loop. Large arrow indicates direction of competitive dominance. a. species D dominates all three of the intransitive loop members. b. species D is dominated by all intransitive loop members. c. species D dominated by one but dominating two others of the intransitive loop members. d. species D dominated by two of the intransitive loop members but dominating the other.*

If the fourth species dominates all three species in the base intransitive loop (fig 4a), that species will eliminate all of the species in the loop. If it is dominated by all three species (fig 4b), it will be unable to invade the community and will go locally extinct. However, if it dominates two of the species in the original loop but is dominated by one of them, a new three-species intransitive loop emerges (B, C, D), with one of the species (A) of the original loop having been eliminated from the community. In this case there are alternative solutions (B,C,D; or A,B,D), depending on initiation conditions. Finally, if the fourth species is dominated by two of the species in the original loop but dominates one of them (fig 4d), either D or B is initially eliminated (depending on initiation conditions). If D is initially eliminated , the original intransitive loop remains. If B is initially eliminated, D is subsequently eliminated followed by A, leaving a monoculture of C. In sum, in none of the cases in figure 5 will the fourth species be incorporated into the system to form a four species permanent community.

It is perhaps not evident why a fourth species cannot enter the system, since the equilibrium values of all three species in the intransitive loop are lower than their carrying capacities (assuming the equilibrium form of the loop), thus reducing some of the competitive pressure on the fourth species. Indeed, it is the case that if the competitive pressure on the fourth species is reduced (but still retaining the basic dominance control between it and the other three species), it is still the case that the fourth species cannot enter into the community permanently. Assuming equation 4, if we allow the extreme situation of B=0, we can compute the equilibrium values of $x_1$, $x_2$ and $x_3$, as:

$$x_i = \frac{1 + A^2 - A}{1 + A^3} \qquad\qquad 7$$

Consider the ODE for species 4 as,

$$\frac{dx_4}{dt} = x_4 \left( \frac{K_4 - x_4 - \alpha_{41} x_1 - \alpha_{42} x_2 - \alpha_{43} x_3}{K_4} \right).$$

If we presume that the dynamics of the intransitive loop are relatively fast, we can presume that the fourth species enters near to the three-species equilibrium. Substitute A for $a_{ij}$, and equation 7 for the equilibria, we can write the equilibrium of the fourth species as,

$$x_4^* = K_4 - 3 \frac{A + A^3 - A^2}{1 + A^3}$$

whence it is evident that $x_4$ will be positive if,

$$K_4 > 3A \frac{1 + A^2 - A}{1 + A^3} \qquad\qquad 8$$

Thus, if equation 8 is true, all four species can coexist. However, it is evident that if equation 8 is true, the assumption that all pairs are dominance controlled is violated. Indeed, it is not possible to add a fourth species to a basic three species intransitive loop to form a stable community of four species (under the assumption of dominance controlled pairs).

There is a simple graphical technique for adjudicating whether the system will be permanent (or have a simple heteroclinic cycle) or decompose, using the graphical representation of figure 5. With reference to figure 6, note that in 6a the simple intransitive triplet is represented. Trivially, each species "effects" a competitive effect on just one other species (the large arrowheads) and must "respond" to the competitive pressure of just one other species. Thus the competitive effect and competitive response are "balanced." Contrarily, in figure 6b, this balance is lost. Species C and B must respond to the effects from two other species whereas species A and D must respond to only one other species, yet exert an effect on two other species. A qualitative glance at figure 6b reveals what can be expected dynamically: B receives pressure from both C and D and is likely to go extinct first, releasing A from any competitive pressure at all, whence C goes extinct from pressure from both D and A, and, finally, A beats D, resulting in a monoculture of A. Furthermore, there is no way of arranging the competitive effects such that all four species will persist, as already discussed in conjunction with figure 5. Adding a fifth species, however, recaptures the possibility of balancing effect and response competition (for each species, the number of large arrowheads pointing to it is equal to the number of small arrowheads pointing to it – fig 6c). When such a balance between effect and response competition occurs, the system as a whole will persist, anchored by one or more intransitive loops. Since the basic intransitivities are responsible for the persistence, it makes sense to talk about this type of structure as an "intransitive structure," whence we conclude that persistence of a subcommunity of more than three species is possible only if the subcommunity comprises an intransitive structure (again, presuming the complete dominance control of all species pairs in the general species pool).

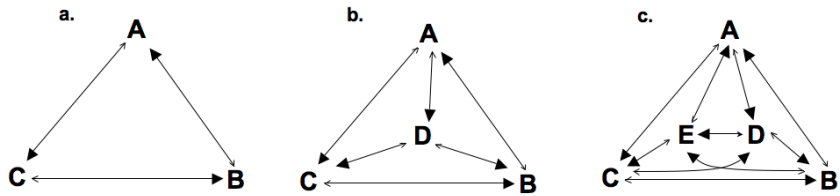

*Figure 6.  Intransitive and transitive structures. a. the simple triplet where the implied balance between effect (large arrowheads) and response (small arrowheads) is balanced (one effect species and one response species per competitor). b. one example of adding a fourth species to the basic intransitive triplet (from figure 5c). Note that balancing effect and response is not possible in this (or any other even) case since an odd number of competitors are involved with each species.  Thus, here, species C and B each have two effect competitors acting on them (species D and C in the case of species B, and species A and D in the case of species C), while species A and D have only one effect competitor but two species to which they must respond.  c. An example of a five species intransitive structure in which each species contends with two effect species and two response species, thus having a perfect balance between effect and response.*

It is important to recall that such intransitive structures (e.g., fig 6c) can be either stable, whence the system will persist in perpetuity, or unstable in which case some form of hetroclinic cycle will emerge, eventually resulting in a single species dominating the system (which of those species will be largely stochastic). From the point of view of the Yodzis categorization, an intransitive structure will behave either as if it were a niche controlled community (if it is a stable intransitive structure) or a dominance controlled community (if it is an unstable intransitive structure).  This general result is surprising when recalling that the underlying structure is assumed to be completely dominance controlled (when evaluating each pair of species separately).

**Dynamic modification of the species pool in a metacommunity**: Initially we assumed that the overall species pool is unchanging, a reasonable assumption in a mainland/island comparison.  However, in light of local dynamics, in any realistic scenario, those local dynamics may have an effect on the composition of the species pool in a metacommunity, perhaps not violating the assumption of a constant species list, but perhaps violating the assumption of a constant relative abundance of each of the of species, which will obviously affect the probability of occupying a particular subhabitat at any point in time. This effect is obviously constrained by the issue of intransitivity.

The issue is qualitatively obvious. If small propagule populations continually occupy subhabitats (as, for example, in the immigration rate of island biogeography) the dynamics of the system will be transient for a significant part of the time. Transient dynamics can be very complicated (Hastings, 2001; 2004; Simonsen et al., 2008), although in simple two dimensional models they appear intuitively obvious (e.g., the classic LV competition equations). By definition, in a metacommunity where the time to equilibrium (which frequently includes local extinction) is not too different from the time to next immigration, transient dynamics are likely to be important. The nature of those dynamics, given the framework of a dominance controlled metacommunity, is qualitatively predictable. Transient dynamics for any dynamical system involve first, an approach to the stable manifold and second, dynamics near that manifold toward the stable equilibrium state (Grebogi, et al., 1983; 1986; Hastings, 2004). In either the two or three dimensional transitive case, the dynamics are relatively simple, with an initial direct approach to the stable manifold and then a direct approach to the equilibrium point on that manifold. The final equilibrium state, which includes extinction of one or two species, is approached relatively rapidly. However, in the intransitive 3-D case the situation is different. If stable, of course, the transient dynamics are oscillatory but converge on a stable state of all intransitive species persistent, as discussed above. If unstable, on the contrary, the transients are extremely important with an initial approach to the stable manifold followed by a heteroclinic cycle which could be very long. Extinction is thus delayed. The consequences for the structure of the overall species pool are dramatic in that those species that could be involved in intransitive loops will tend to become more common in the community and species not typically involved in any intransitive loops will tend towards extinction.

More explicitly, suppose in any given subhabitat species arrive at a rate m, where m is a small number of individuals over some critical time period, $\tau$. The first species to arrive (species A) will be unencumbered by competition and rapidly approach its carrying capacity.  If we presume that something less than 100% attainment of K occurs within $\tau$, a second species (species B) can be expected to arrive before the first species has theoretically attained its full K.  By the assumption of dominance control, either species A or B will tend, first to the stable manifold and second to its K while the other species will tend toward 0. Again assuming less than 100% attainment of these equilibrium values occurs within $\tau$, the third species (species C) will arrive before those equilibrium values occur. If the triplet, ABC, is transitive, the general approach to K,0,0 (or 0,0,K or 0,K,0) will proceed unabated. However, if ABC is intransitive, first the system will approach the stable manifold, and then to either a focal point equilibrium where all three species are significantly greater than zero or a heteroclinic cycle wherein each of the carrying capacities is sequentially approached.  In either case, the loss of all individuals of two of the three species expected in the transitive case is reversed and within some time frame all three species will contribute individuals to the species pool.  The basic idea is illustrated in figure 7.  The relationship between time to colonization ($\tau$) and time to extinction is a complicated one and beyond the intended scope of this article.

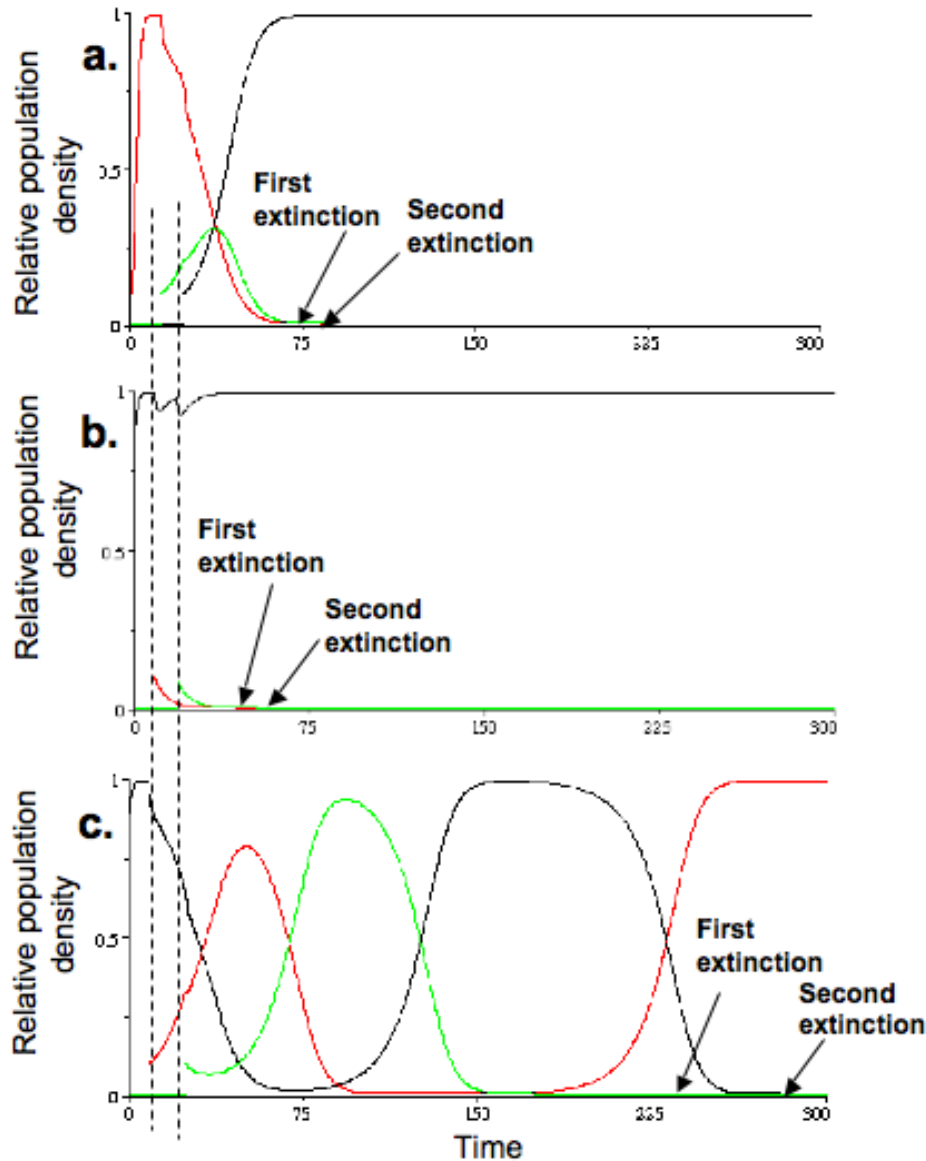

*Figure 7. Time series for exemplary subhabitat for transitive (a and b) versus intransitive (c) triplet, for unstable dominance-controlled pairs. Vertical dashed line indicates time of migration. Note the dramatic difference in time to extinction for transitive versus intransitive. The stable dominance-controlled situation is similar, but more exaggerated since time to extinction for the intransitive case is infinite by definition.*

As mentioned previously, if the intransitive loop is stable, the dynamics are obvious. An illustration is presented in figure 8. In this example, the transitive chain (species A, B, C, D) is augmented at its lowest level by an intransitive loop (C,D,E). In the end the overall structure indicates species C will dominate numerically (it sometimes will dominate because it is the best competitor in the transitive chain, and sometimes will co-dominate with the other two members of the intransitive loop). Over the long run the subdominant species in the transitive chain (A and B) will exist in the system as fugitive species, each forming a separate metapopulation if their migration rate is greater than the extinction due to competition from the other species (which is true in this artificial example).

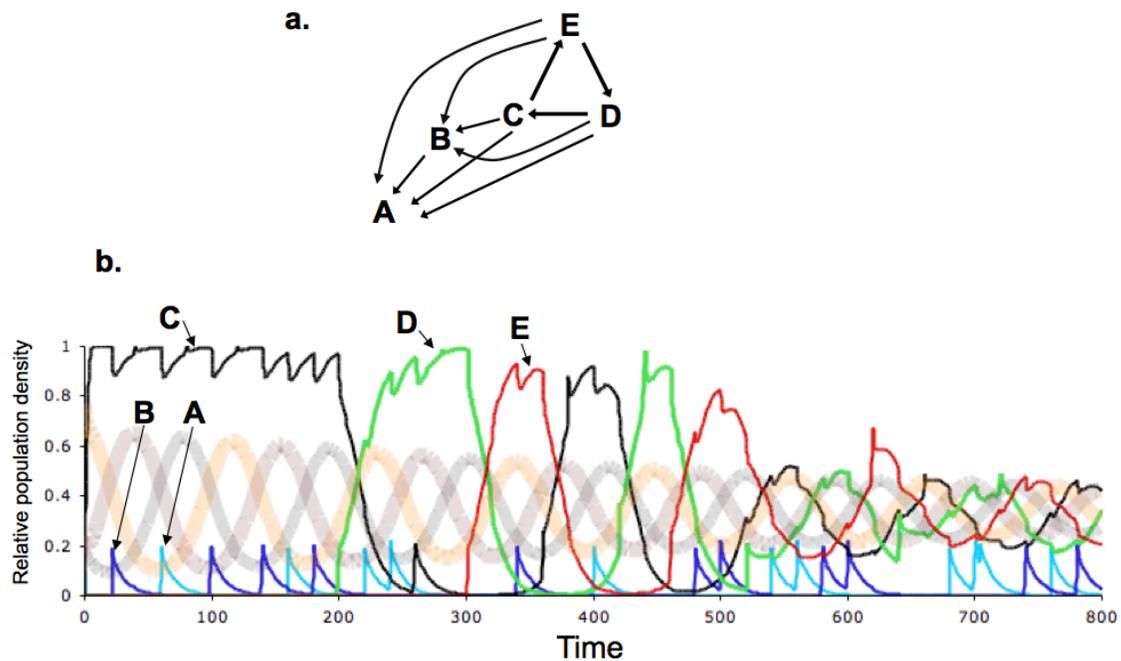

*Figure 8. Illustration of intransitive loop long-term dominance. a. structure of the species pool in this example, where arrowhead indicates "beats" in competition (i.e. B beats A; E beats D). Note, E, C, D form an intransitive triplet whereas all other triplets are hierarchies. b. time series with 0.2 fraction of K introduced at 20 time unit intervals (modeling regular migration from the species pool). Once all three species of the intransitive loop (C,D,E) have migrated (at about time = 300), those species together resist the invasion of A and B, reducing the later two species to the status of fugitive species. Background transparent trajectories are simulations of the intransitive loop alone, with no immigration from the species pool.*

The consequences of this transitive/intransitive dynamics at the local level suggests that those individuals involved in intransitive triplets in the species pool generally, will leave more individuals in that species pool over time, compared to those species involved in only transitive triplets in the species pool. Species that are dominant in a transitive chain as well as members of an intransitive loop will become especially abundant in the species pool, while the subdominant species in the transitive chain will reach the level of fugitive species, as noted above.

**Discussion**: It is perhaps surprising that a three-species subcommunity sampled from a species pool composed of nothing but dominance controlled species pairs can result in a stable three species situation, as argued above. Yet a perusal of the literature finds that intransitive networks may not be all that uncommon amongst competitors (e.g., Buss, 1980; Lankau et al., 2011; Kerr et al., 2002; Kirkup et al., 2004) even though active searching for such arrangements has not been common. Especially among plant communities where transitive networks are generally thought to be common (Keddy and Shipley, 1989), there seems to be an assumption that such transitivities imply that competition is simply strong (which implies dominance or founder control) and locally we expect competitive exclusion almost always. This bias is perhaps partly a function of the nature of the organisms under study, where most species are of similar stature and niche requirements.

The results in the present study apply to the "frictionless" theoretical system of multiple species in competition with one another. Connections to the natural world are obviously tentative. Indeed, I know of no study that attempts to specifically link the ideas herein to metacommunities. Nevertheless, there are countless studies suggesting that the question of both dominance control and intransitive loops are generally important, implying that the question need not remain obscure in the future.

In particular, Keddy and Shipley (1989) review a number of studies in plant ecology both from the point of view of what they refer to as asymmetry (corresponding to the use here of dominance) and transitive pathways (including three species pathways). Although their major conclusion is that asymmetry is almost universal among similarly statured plants and almost all pathways are transitive, it is clear from their analysis that intransitivities also exist, even if not common. The use of DeWitt replacement series in this particular study has been criticized, suggesting that the paucity of intransitivities may in fact be spurious (Herben and Krahulec, 1990).

The expectation that 25% of randomly sampled triplets will be intransitive, as noted above, is derived from the assumption that the competition coefficients in the species pool are all selected at random. The natural question that arises is whether the species pool in general can be thought of as allocating competition at random. It is not common that experimental procedures are adequate for any given community to assess the competition coefficients with regard to intransitivity (Herben and Krahulec, 1990). One example is the competition experiments with seven species of prairie plants by Goldberg and Landa (1991), in which target plants were subjected to competitive effects from a variable number of competitors, although even in this case it is presumed that response to competition is constant with respect to the initial biomass of the target species. From these seven species there are 35 possible distinct triplets, and the theory suggests that we would expect nine of them to be intransitive if the allocation of competition coefficients is at random. In fact, only two were (*Trifolium repens, Chenopodium album, Phleum pratense; Amaranthus retroflexus, Chenopodium album, Phleum pratense*), based on the per target plant results (table 2a in Goldberg and Landa, 1991). All species in this study were of similar stature and likely with similar basic requirements. It is also the case that the species were not necessarily native nor clearly part of the same natural community, so the expectation that intransitive triplets would increase in the species pool over time may not apply.

Buss and Jackson (1989) note that several studies of corals find strict transitivity (Lang, 1971,1973), but then report on a system of seven species of Jamacian cryptic reef species. Of the possible 35 triplets, at least five of them are intransitive, well below the expected eight or nine from the random expectation, but clearly a deviation from the frequently assumed transitive universality.

Although randomly assigning competition coefficients to species will inevitably result in some intransitive triplets (25% of all possible triplets), the question remains for any particular triplet what is the mechanism of competition that could result in the intransitive structure. Indeed a variety of well known competitive mechanisms could result in such structures. Consider, for example, the general consensus regarding the competitive structure of ant communities (Holdober and Wilson, 1990; Par and Gibb, 2010). This structure is commonly thought to be a result of two or sometimes three foraging strategies, the "discoverers" (those species that tend to discover a resource rapidly), the "dominators" (those species that take over larger resources and prevent other species from using them) and the "insinuators" (those species that avoid interactions with other species and take advantage of resource acquisition actions not immediately available to other species).  As noted elsewhere (Perfecto , 1994; Perfecto and Vandermeer, 2013) discoverers will be favored when resources are generally small while dominators will be favored when resources are generally large.  Thus the process of competition in ants, sometimes thought of as a balance between discovery and dominance, can also be thought of as operating in the space of two dimensions (large resources versus small resources), in which case three species with the three distinct foraging strategies could easily be functioning in an intransitive loop (Vandermeer 2011), as suggested elsewhere (Perfecto and Vandermeer 2013), where the discoverer species reduces the abundance of the small resource resulting in an environment biased with large resources and thus favoring the dominator species.  Between the two extremes (mainly small resources versus mainly large resources) the extirpator finds temporary advantage over both dominator and discoverer, leading to potential persistence of the three, as shown diagrammatically in figure 9a and b

.

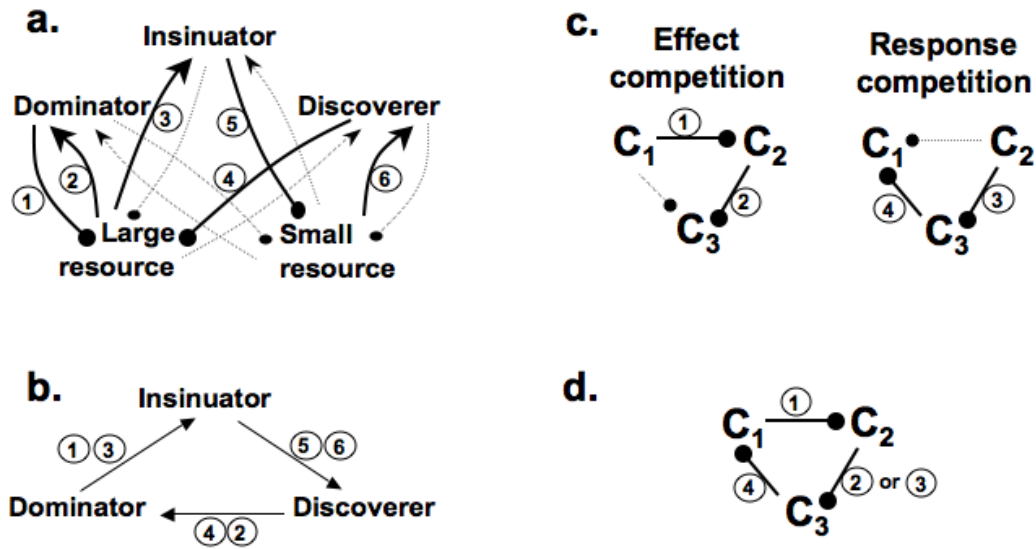

*Figure 9. Two potential mechanisms for generating intransitive loops in the context of basic dominance control (arrowheads here represent positive effects and closed circles represent negative effects). a. arrangement common in ant competition where three species engage two resource types with different foraging strategies (see text). If the environment contains only small resources, the insinuator beats both the discoverer and the dominator while the discoverer beats the dominator. If the environment contains only large resources, the dominator beats the insinuator and the discoverer and the insinuator beat the discoverer. Thus with either all large resources or all small resources the three species are completely transitive. b. the intransitive triplet that results when the bold numbered interactions are much larger than the dotted interactions, in a. c. possible arrangement in plant competition where response competition and effect competition have different patterns of dominance hierarchy. d. the intransitive triplet that results when the bold numbered interactions are much larger than the dotted interactions.*

There are other potential intransitive mechanisms in competition. Any time there are two niches, the potential for distinct complementarity on the two distinct niche axes creates an obvious potential for intransitivity (Vandermeer, 2011). In the context of plant competition, the well-known distinction between effect competition and response competition (Goldberg and Landa, 1991) creates the potential for an intransitive triplet (fig. 9c and d). Many other examples could be cited.

The general qualitative analysis of systems in which all species pairs are dominance controlled, suggests that the only multiple species sub-communities that can persist over the long term are those that are structured based on one or more intransitive loop, a condition that might be referred to as an intransitive structure. If it is true, as some have suggested (Keddy and Shipley, 1989; Goldberg and Landa, 1991) that any pair of species in a large competitive community will be characterized by dominance in competition, it is also a reasonable hypothesis to suggest that any large competitive community must have an intransitive structure, as implicitly suggested by Buss and Jackson (1989) and theoretically advanced elsewhere (Vandermeer and Yitbarek, 2012).  The hypothesis of intransitive structure should be considered as an alternative to other hypotheses of coexistence in competitive communities (Hubbell, 2001; Chesson, 2000; Bever, 2003; Amarasekare, 2003; Chase, 2005).

In any large community structured by competition, if there is a super structure that implies either a mainland/island community or a metacommunity, with subcommunities repeatedly drawn from the species pool, and if all species pairs in the species pool are dominance determined (as is implied by the general idea of transitivity, or dominance hierarchy, to start with), it is unlikely that intransitive triplets will not exist somewhere within that species pool. If such is the case, the repeated operation of invasion and extinction will sometimes result in larger local communities, anchored to one or more intransitive triplets, a structure referred to herein as an intranstitive structure.  An additional dynamic process is likely to operate specifically in the case of a metacommunity, in which the overall species pool may be modified in the direction of those species having a tendency to form intransitive substructures.